\documentclass[12pt,a4paper]{scrartcl}

\usepackage{ILD}
\usepackage{arydshln}
\usepackage[bottom,perpage,symbol*]{footmisc}
\usepackage{feynmf}
\usepackage{textpos}
\usepackage{lineno}
\usepackage{enumitem}
\usepackage{verbatim}
\usepackage{csquotes}
\usepackage{fancyhdr}
\usepackage{flushend}
\usepackage{authblk}
\usepackage{amsmath}
\usepackage{amssymb}
\usepackage{upgreek}
\usepackage{titling}

\usepackage{graphicx}
\usepackage{placeins}
\usepackage{float}
\usepackage{caption}
\usepackage{subcaption}
\usepackage{color}
\usepackage{url}
\usepackage{hyperref}

\title{Two-particle angular correlations in the search for new physics at future $e^{+}e^{-}$ colliders}

\titlecomment{Talk presented at the International Workshop on Future Linear Colliders (LCWS 2023), 15-19 May 2023. C23-05-15.3.}

\date{\today}
\draft{LCWS2023}{v1}

\addauthor{E.~Musumeci}{\institute{1} \footnote{Speaker \\ \href{mailto:emanuela.musumeci@ific.uv.es}{emanuela.musumeci@ific.uv.es} \\ \href{mailto:emanuela.musumeci@cern.ch}{emanuela.musumeci@cern.ch}}}
\addauthor{R.~Perez-Ramos}{\institute{2}$^{,}$\institute{3}}
\addauthor{A.~Irles}{\institute{1}} 
\addauthor{I.~Corredoira}{\institute{4}}
\addauthor{V.~A.~Mitsou}{\institute{1}}
\addauthor{E.~Sarkisyan-Grinbaum}{\institute{5}$^{,}$\institute{6}}

\addauthor{M.~A.~Sanchis-Lozano}{\institute{1}}

\addinstitute{1}{IFIC, Universitat de Val\`encia and CSIC, C/ Catedr\'atico Jos\'e Beltr\'an 2, \\ E-46980 Paterna (Valencia), Spain\\}

\addinstitute{2}{DRII-IPSA, 63 bis Boulevard de Brandebourg, 94200 Ivry-sur-Seine, France, and\\}

\addinstitute{3}{Laboratoire de Physique Th\'eorique et Hautes Energies (LPTHE), UMR 7589,\\ Sorbonne Universit\'e et CNRS, 4 place Jussieu, 75252 Paris Cedex 05, France\\}

\addinstitute{4}{Instituto Galego de F\'isica de Altas Enerx\`ias (IGFAE, USC), Spain\\}

\addinstitute{5}{Experimental Physics Department, CERN, 1211 Geneva 23, Switzerland, and\\}

\addinstitute{6}{Department of Physics, The University of Texas at Arlington, \\ Arlington, TX 76019, USA\\}

\abstract{
  The analysis of angular particle correlations can yield valuable insights into the initial state of matter
in high-energy collisions, thereby potentially revealing the existence of Beyond the Standard Model scenarios such as Hidden Valley (HV). In this study, we focus on a
QCD-like hidden sector with relatively massive HV quarks ($\lesssim 100$~GeV)
which might enlarge and strengthen azimuthal correlations of final-state SM
hadrons. In particular, we study the formation 
and possible observation of \textit{ridge-like} structures in the angular two-particle correlation function at future 
$e^+e^-$ colliders, with a much cleaner environment than in hadron colliders, such as the LHC.}

\graphicspath{ {./logos/}{./figures/} }

\begin{document}

\titlepage



\newpage

\section{Introduction}
\label{sec:intro}

Correlations have traditionally played and still play a fundamental role in many branches of physics, from quantum mechanics to nuclear and elementary particle physics, astrophysics and cosmology. In particular, the study of two-particle angular correlations has proved to be an efficient method to investigate the underlying mechanisms of (multi)particle production in high-energy collisions in cosmic and accelerator physics, potentially uncovering collective effects of matter at extreme conditions of temperature and density.

Indeed, angular correlations have nowadays become a fundamental observable in heavy-ion collision studies. Moreover, the discovery of a long-range near-side ridge also in $pp$ collisions~\cite{CMS_2015_ridge_pp}, together with the characterisation of this effect in different collision systems and pseudorapidity ranges, has become crucial for its  interpretation. However, a definitive explanation of this phenomenon remains under discussion (controversial) as various theoretical explanations have emerged. On the other hand, let us point out that no clear ridge-like signal appeared in the $e^+ e^-$ collision data analysed by ALEPH~\cite{Badea_2019} and BELLE~\cite{Chen_2022} collaborations, the former reaching centre-of-mass energy up to $209\;\mathrm{GeV}$.

This work considers two-particle angular correlations as a possible tool to look for physics beyond the Standard Model (BSM) at future $e^+ e^-$ colliders, which should run at higher energies. In particular, we focus on the so-called \textit{Hidden Valley} (HV) scenario, which stands for a wide class of models with one or more hidden sectors beyond the Standard Model (SM), with not necessarily too heavy new (valley) particles.

\section{Two-particle angular correlations}
\label{sec:2pac}
The thrust reference frame will be used throughout this work, so the rapidity $y$ of a particle is always referred to the thrust coordinate axis. At the same time, the azimuthal angle $\phi$ is calculated with respect to the
eigenvector of the thrust tensor having the smallest eigenvalue, in the plane perpendicular to the thrust axis event by event. Note that only rapidity and azimuthal differences of final-state SM particles 1 and 2 ($\Delta y= y_1-y_2$, $\Delta \phi= \phi_1-\phi_2$) are the main variables involved in this observable.
The two-particle correlation function can be defined as a ratio of the signal over the background as shown in Eq.\;(\ref{eq:corr_function}),

\begin{equation}
    C^{(2)}(\Delta y,\; \Delta \phi) = \frac{S(\Delta y,\; \Delta \phi)}{B(\Delta y,\; \Delta \phi)}.
\label{eq:corr_function}
\end{equation}

By definition, the signal $ S(\Delta y, \Delta \phi) $ provides the density of particle pairs within the same event, while the background, $ B(\Delta y, \Delta \phi) $, corresponds to the density of pairs from mixed events. Equations\;(\ref{eq:same_evt})--(\ref{eq:mixed_evt}) stem from a sample of tracks coming from randomly chosen events,

\begin{equation}
    S(\Delta y,\; \Delta \phi) = \frac{1}{N_\text{pairs}} \frac{d^{2}N^\text{same}}{d \Delta y d \Delta \phi},
\label{eq:same_evt}
\end{equation}

\begin{equation}
    B(\Delta y,\; \Delta \phi) = \frac{1}{N_\text{mix}} \frac{d^{2}N^\text{mix}}{d \Delta y d \Delta \phi}.
\label{eq:mixed_evt}
\end{equation}

Let us note that following the above definition, no correlation between particles implies $C(\Delta y,\; \Delta \phi) = 1$. Positive (negative) correlations correspond to values higher (smaller) than unity.

Focusing on the azimuthal part of the correlation function, the so-called azimuthal yield, $Y(\Delta \phi)$, is defined by integrating over a $\Delta y$ range as shown in Eq.\;(\ref{eq:yield}). Such an integration range is suitable to avoid correlations among particles within the same jet, whose physical interest lies outside the scope of this work,

\begin{equation}
    Y(\Delta \phi) = \frac{\int_{  1.6 \leq |\Delta y| \leq 3 } S(\Delta y,\; \Delta \phi) dy}{\int_{1.6 \leq |\Delta y| \leq 3 } B(\Delta y,\; \Delta \phi) dy}.
\label{eq:yield}
\end{equation}

\section{Hidden Valley Phenomenology}
\label{sec:hv_pheno}
In most HV models, the SM gauge group sector $G_{SM}$ is extended by (at least) a new gauge
group $G_V$ under which all SM particles are neutral. Moreover, there is a new category of particles, called \textit{v-particles} charged under $G_V$ but neutral under $G_{SM}$. On the other hand, communicators charged under both $G_{SM}$ and $G_{V}$ are introduced in the theory to allow interactions between SM and v-particles. 
The effects of the hidden sector on the visible sector depend on the way the former communicates with the SM~\cite{Carloni_2010}. In this work, we are especially interested in the effects of the invisible (i.e. HV) sector
on the partonic cascade into final-state SM particles in
high-energy collisions.

Hereafter, we will consider a QCD-like HV scenario including the corresponding $q_v$-quarks and $g_v$-gluons together with a strong coupling constant $\alpha_v$ that, for simplicity, is assumed not to run but to be fixed at a given value  $\alpha_v = 0.1$.
The communicators (generically denoted as $F_v$) are considered mirror partners of SM-charged quarks and leptons, and they have to be pair-produced. For some values of the HV parameters, such particles can promptly decay into a particle $f$ of the visible sector (the SM partner) and one $q_v$ of the hidden sector according to the splitting:  $F_v \to f q_v$.  In $e^+ e^-$ collisions, these particles can be produced via the process $e^+ e^- \to \gamma^* /Z \to F_v\Bar{F}_v \to$ hadrons. As we shall see, the $q_v$ mass strongly influences the kinematics of the visible cascade leading to SM particles. This mass is currently unconstrained and can then range from zero to close to $F_v$ masses~\cite{PhysRevD.105.053001}.\newline

\section{Preliminary Results: Signal vs.\ Background}
\label{sec:results}
\begin{figure}[!h]
\centering
\hspace{.8cm}
\begin{subfigure}{0.43\textwidth}
\includegraphics[width=\linewidth]{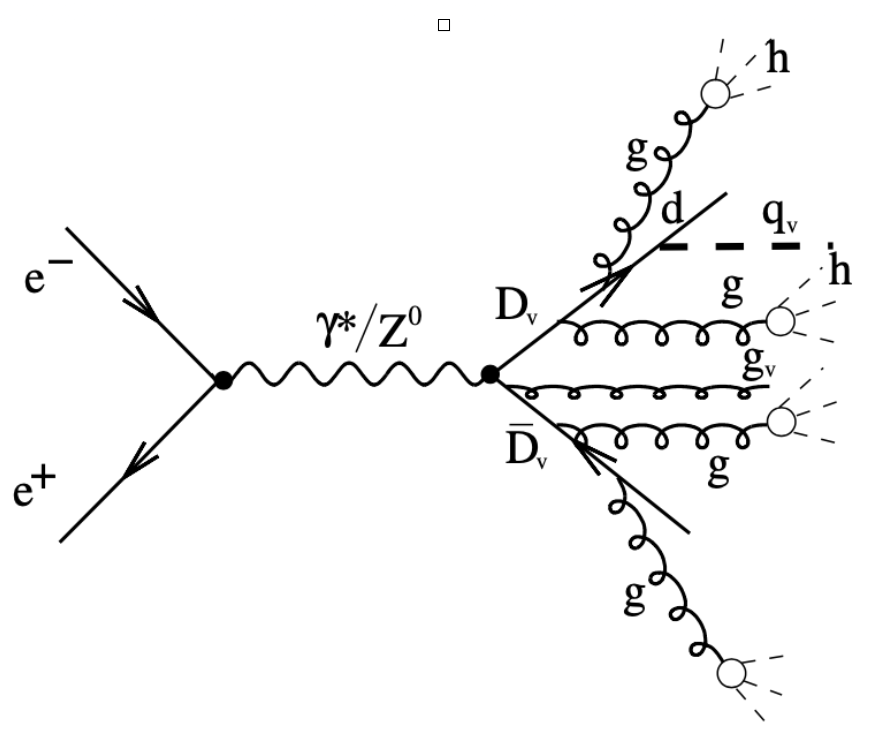}\caption{HV Signal} \label{fig:1a}
\end{subfigure}
\hspace{.8cm}
 \begin{subfigure}{0.43\textwidth}
\includegraphics[width=\linewidth]{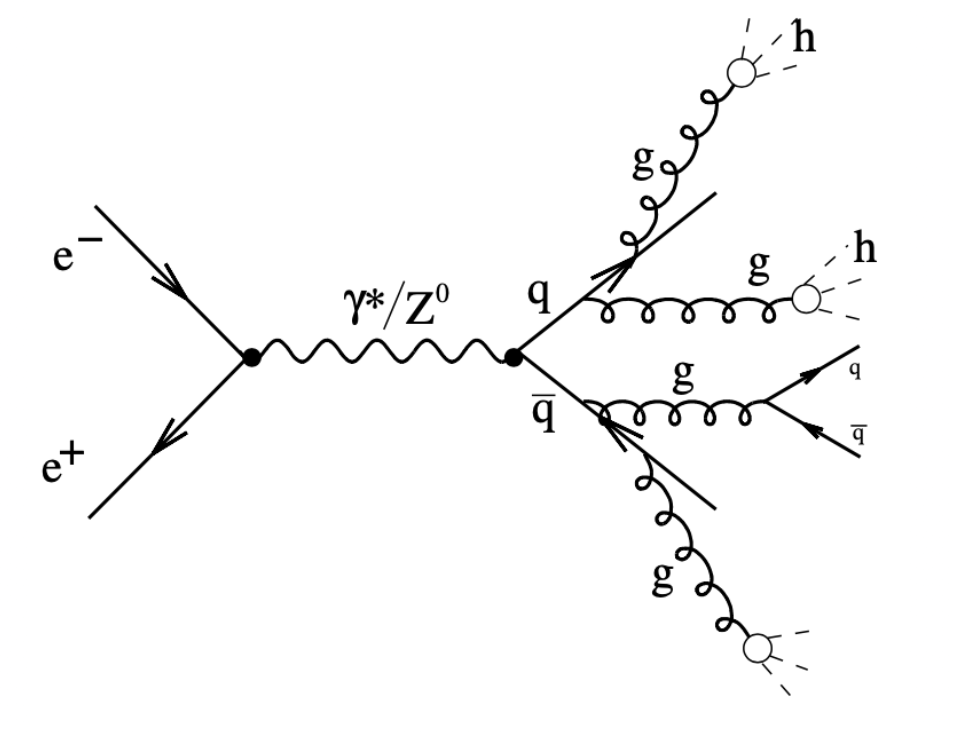}
\caption{ SM Background} \label{fig:1b}
\end{subfigure}
\hspace{.8cm}
\caption{\footnotesize \centering Leading diagrams of Hidden Valley \subref{fig:1a} and SM light quarks \subref{fig:1b}, up to the bottom $b$, production in $e^+ e^-$ collisions at $\sqrt{s}=250\;\mathrm{GeV}$. } \label{fig:1}
\end{figure}

To start the study of potentially observable signals in $e^+e^-$ collisions due to HV radiation along the partonic cascade, the following tools were used at the particle level:
\begin{itemize}
    \item PYTHIA8~\cite{Sjostrand_2015} as Monte Carlo event generator;
    \item FastJet~\cite{Cacciari_2012} for jet-clustering. In particular, the algorithm considered is the generalised $k_t$ algorithm for $e^+ e^-$ collisions, i.e. \textit{ee$\_$genkt$\_$algorithm};
    \item ROOT to perform the analysis~\cite{citeulike:363715}.
\end{itemize}
The centre-of-mass energy $\sqrt{s}$ of $e^+e^-$ collisions was set to $250\;\mathrm{GeV}$ in order to coincide with the expected first commissioning of the collider as a Higgs boson factory. The HV signal considered in this work is based on the process $e^{+}e^{-}\to D_v\bar{D}_v\to$ hadrons, as depicted in Fig.~\ref{fig:1a}, where the heavy v-quark $D_V$ is assumed to be the mirror partner of the SM $d$-quark. In our study, we set as a benchmark scenario $\alpha_{v}=0.1$, $m_{D_v}=125\;\mathrm{GeV}$, and three different mass values for the $q_v$-quark, namely, $m_{q_v}= 10 \;\mathrm{GeV}$,  $m_{q_v}= 50 \;\mathrm{GeV}$ and  $m_{q_v}= 100 \;\mathrm{GeV}$. Higher $\sqrt{s}$ values and more scenarios will be explored in a forthcoming paper.

The decay modes $g_v \to g_v \Bar{g}_v$, $g_v \to q_v\Bar{q}_v$ and $\gamma_v \to q_v \Bar{q}_v$ are essentially similar to SM processes along the partonic cascade before hadronisation~\cite{PhysRevD.105.053001}.  
Let us stress that in our analysis, the background consists of the inclusive production of all the SM quark species except the top flavour (since $\sqrt{s}=250\;\mathrm{GeV}$  lies below its production threshold of about $\sqrt{s}=350\;\mathrm{GeV}$), namely, $e^{+}e^{-} \to q \bar{q}\to$ hadrons, where $n_f=5$ for $q=u,d,s,c,b$ (Fig.~\ref{fig:1b}) and $q$ collectively denotes the ``lightest'' quark flavours up to the bottom quark. Additional backgrounds, e.g., due to Initial State Radiation or Vector Boson Fusion, will be analysed in a future paper, where  detector effects will also be incorporated.

Following from PYTHIA generation, the cross-sections for the HV and SM processes are reported in Table~\ref{tab:cross_sections}.

\begin{table}[!h]
\centering
\begin{tabular}{ |p{5.7cm}|p{3.5cm}|  }
\hline
\centering Process &  $\sigma_\text{PYTHIA8} (\;\mathrm{pb})$ \\
\hline
$e^{+}e^{-}\to D_v\bar{D}_v$, $m_{q_v}=10\;\mathrm{GeV}$ & $1.243 \cdot 10^{-1}$ \\
$e^{+}e^{-}\to D_v\bar{D}_v$ , $m_{q_v}=50\;\mathrm{GeV}$ & $1.244 \cdot 10^{-1}$ \\
$e^{+}e^{-}\to D_v\bar{D}_v$,  $m_{q_v}=100\;\mathrm{GeV}$ & $1.199 \cdot 10^{-1}$  \\
$e^{+}e^{-}\to q\bar{q}$ & $1.220 \cdot 10^{1}$ \\
\hline
\end{tabular} 
     \caption{ \centering \footnotesize Cross section values of $e^{+}e^{-}\to D_v\bar{D}_v$ processes for three different masses $m_{q_v}$ and $e^{+}e^{-} \to q \bar{q}$ at $\sqrt{s}=250\;\mathrm{GeV}$  }
    \label{tab:cross_sections}
\end{table}
Due to some foreseen detector acceptance conditions and in order to account for  realistic event reconstruction performance, several selection cuts were set on generated events: (i)   
we select final-state particles with transverse momentum $p_{\mathrm{T}}>0.5$~GeV with respect to the beam axis, (ii) $|\cos{\theta}|\leq 0.99$ due to the detector acceptance,  where $\theta$ is the angle with respect to the beam axis as well. These basic cuts are motivated by expected Higgs Factories detector performance studies reported in Ref.~\cite{ILDConceptGroup:2020sfq}. 

\begin{figure}
\centering
\includegraphics[width=0.55\linewidth]{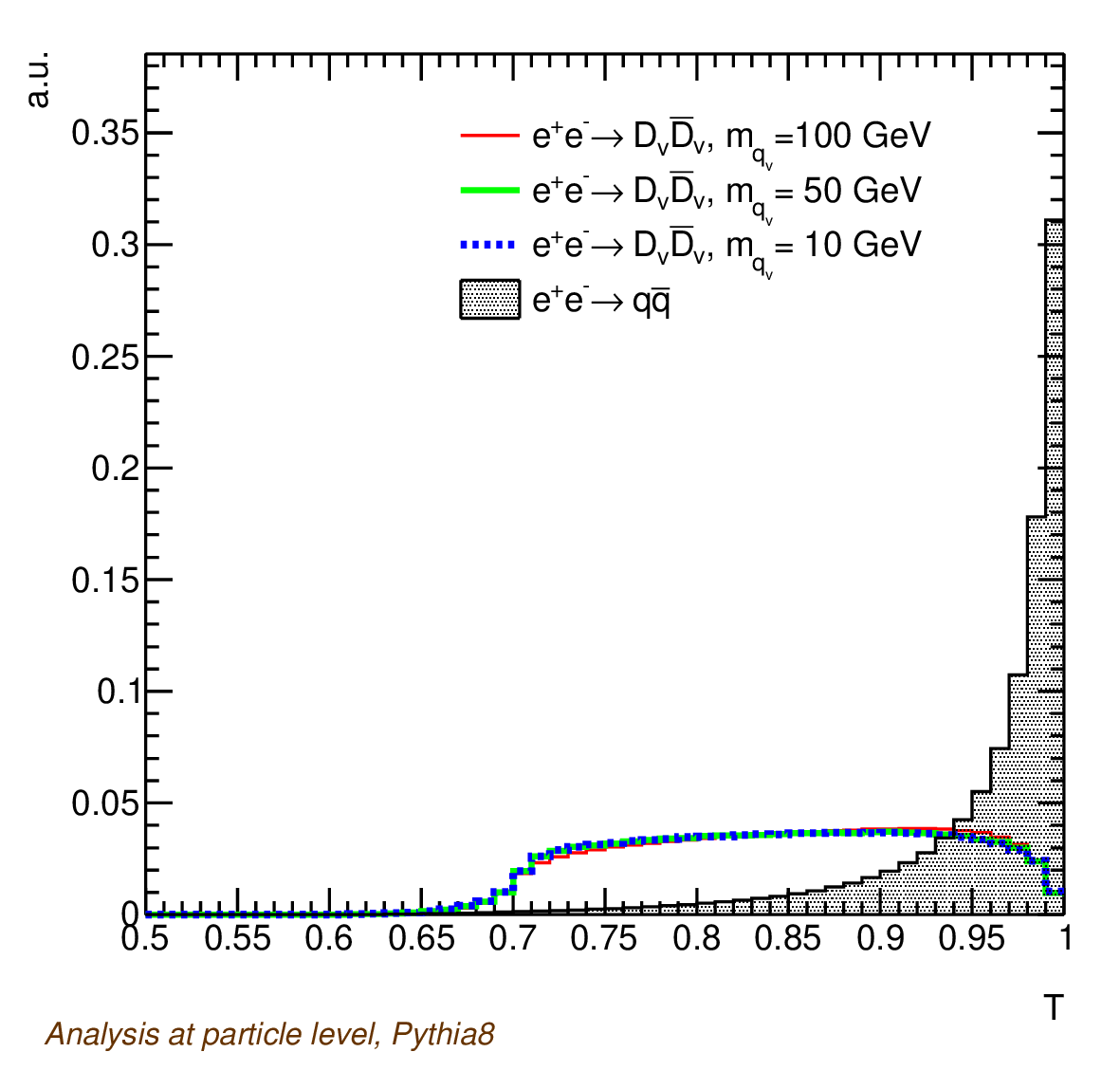}
\caption{\footnotesize \centering Thrust distribution for the HV signal considering different v-quark masses $m_{q_v}$ and SM process (shaded), normalised to unity.} \label{fig:thrust}
\end{figure}

In our analysis, we work in the thrust reference frame,
where the thrust ($T$) of an event is defined as
\begin{equation}
    T= \max_{\vec{n}} \frac {\sum_i |\vec{p}_i \cdot \vec{n}|}{\sum_i |\vec{p}_i|},
\end{equation}
where $|\vec{n}|= 1$ and the sum runs over the three-momenta of all final-state particles, event by event. 
The thrust axis is defined by the eigenvector $|\vec{n}_T|$ of the thrust tensor $T$ having the largest eigenvalue. This definition
means that for $T = 1$, the event is perfectly back-to-back and pencil-like, while for $T \approx 1/2$, the event is spherically symmetric. Notice that, as depicted in Fig.~\ref{fig:thrust}, light quark production leads mostly to pencil-like events, whilst HV production yields sphere-like events on average.

\begin{figure}
\centering
\includegraphics[width=0.55\linewidth]{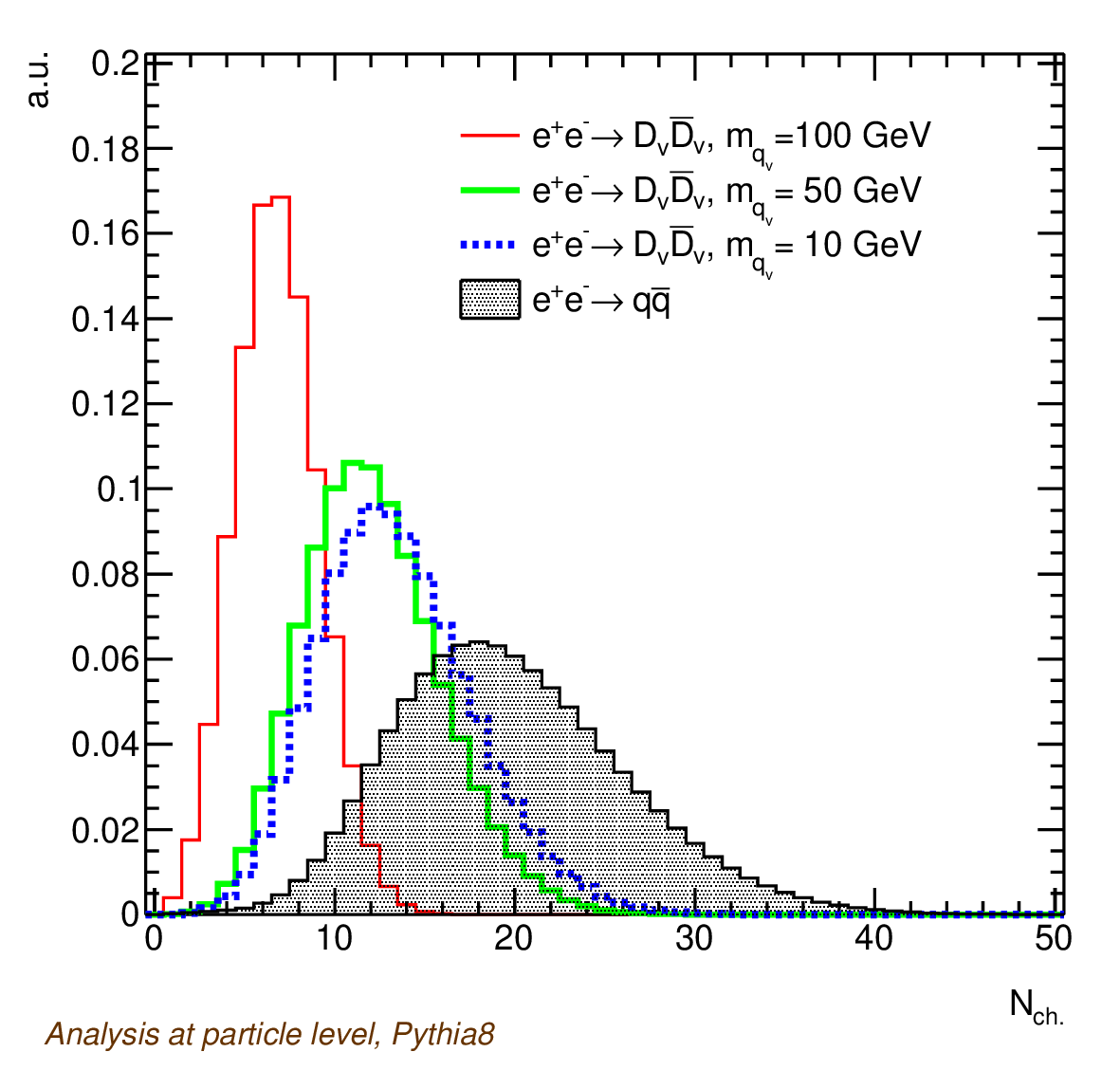}
\caption{\footnotesize \centering Multiplicity of charged particles for the HV production with three different $q_v$-quark masses $m_{q_v}$ and the SM process (shaded). The distributions are normalised to unity. } \label{fig:multiplicity}
\end{figure} 

\begin{figure}
\centering
\includegraphics[width=0.55\linewidth]{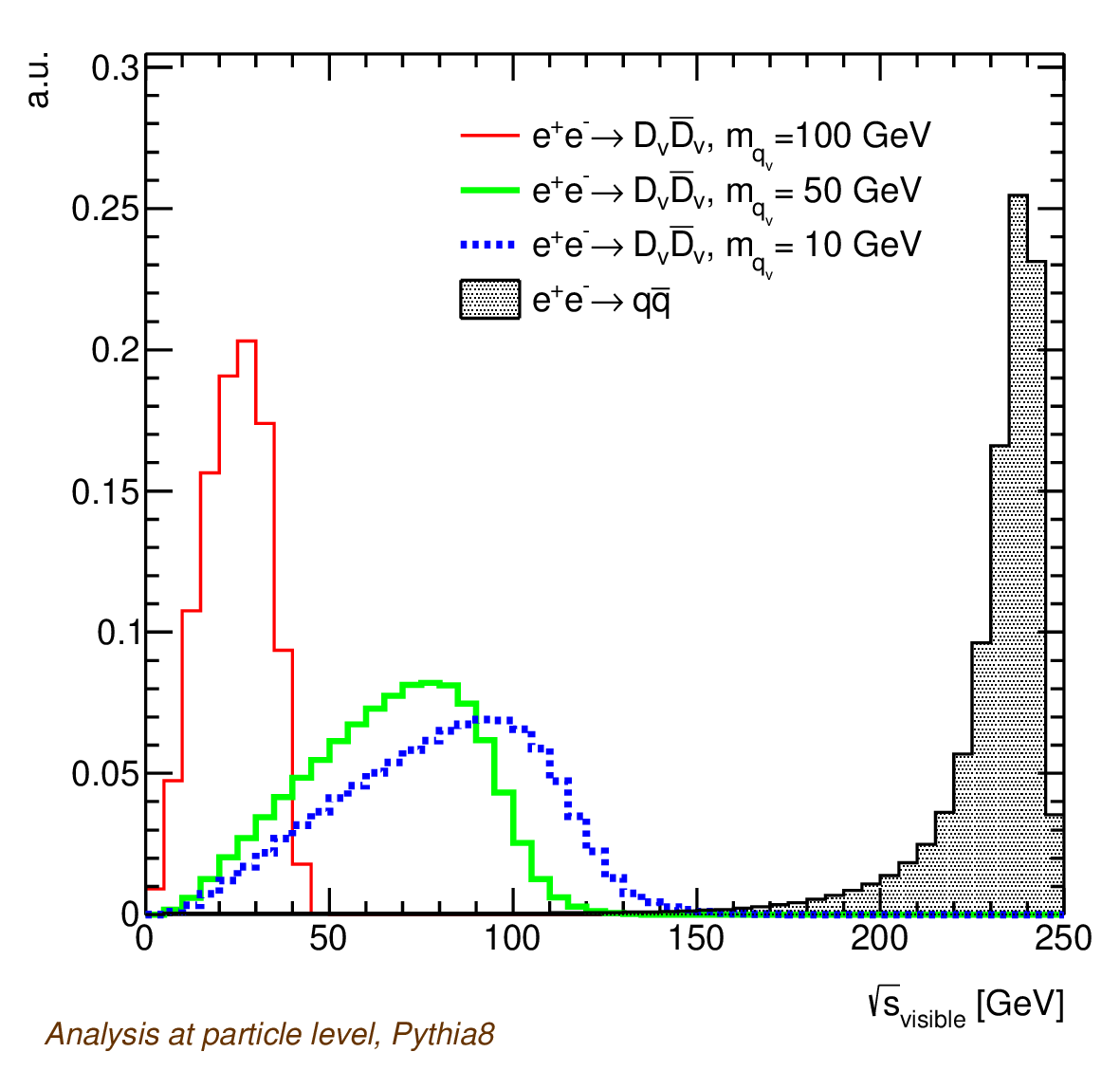}
\caption{\footnotesize \centering Di-jet invariant mass distribution for HV production with different v-quark masses $m_{q_v}$, and the SM process (shaded). The distributions are normalised to unity. } \label{fig:masa_invariante}
\end{figure} 

The distributions of the charged particle multiplicity, thrust and di-jet invariant mass for either HV signal and $q\bar{q}$ production are presented in Figures~\ref{fig:multiplicity},~\ref{fig:thrust} and~\ref{fig:masa_invariante}, respectively. The normalisation factor is set equal to unity, without considering the cross-section values for signal and background. Indeed, on the $y$-axis, ``a.u.'' stands for ``arbitrary units'', as the main focus here is to study the different shapes and behaviours that can allow one to distinguish between signal and background. Nevertheless, given the significant difference between the
values of the cross-section of the signal and background,  it is necessary to remove as much background as possible. Therefore, a selection
cut based on the di-jet invariant mass variable was applied to all generated events. By inspection of the invariant mass distribution (Fig.~\ref{fig:masa_invariante}), it turns out that it is possible to separate the HV signal from the background by applying a cut on such a kinematic variable. The specific cut to be applied actually depends on the $q_v$ mass. 

\begin{figure}
\begin{subfigure}{0.5\textwidth}
\includegraphics[width=\linewidth]{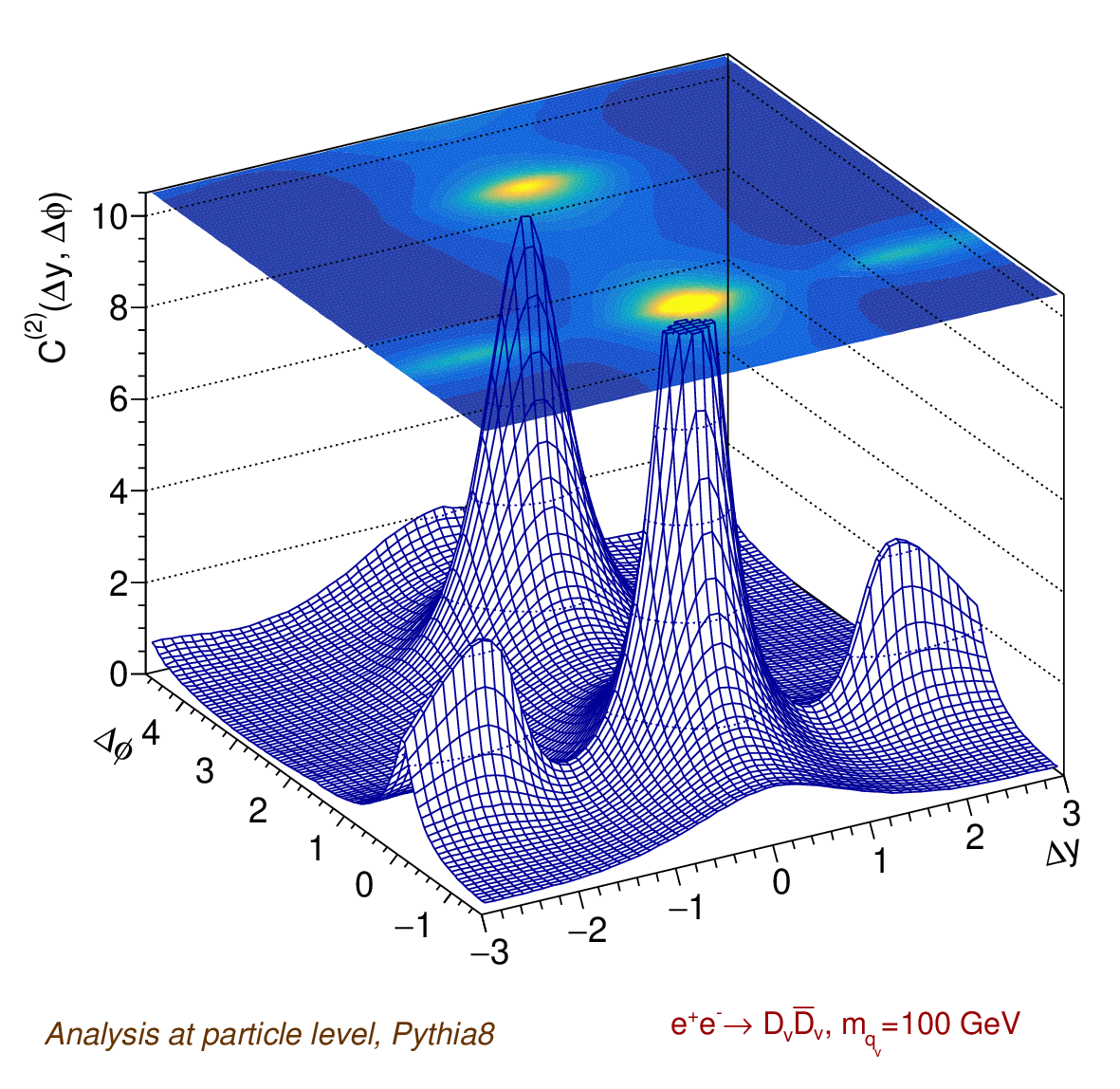}
\caption{\footnotesize \centering Two-particle angular correlation function for the pure HV Signal} \label{fig:corr_hv}
\end{subfigure}
\begin{subfigure}{0.5\textwidth}
\includegraphics[width=\linewidth]{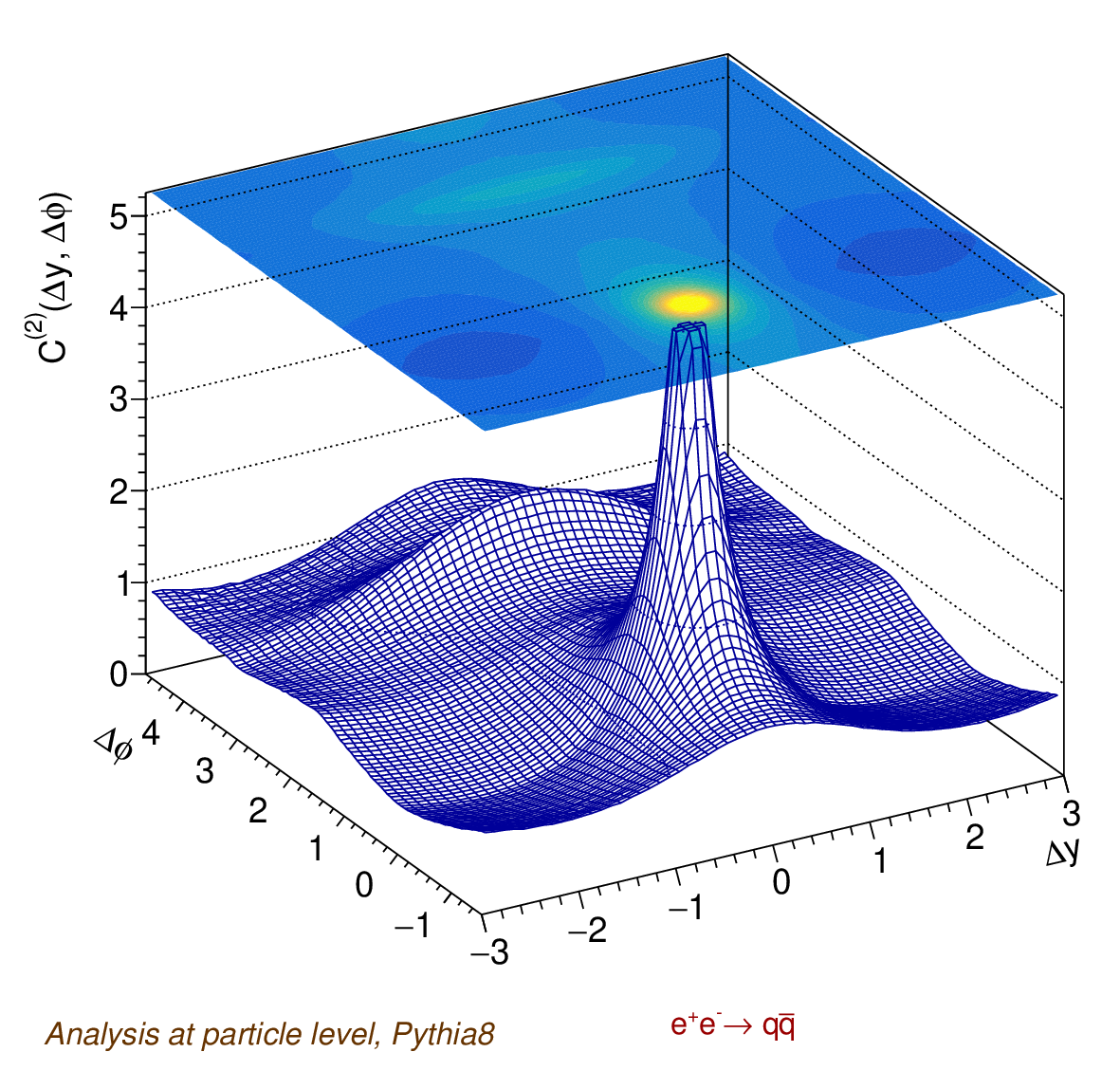}
\caption{\footnotesize \centering Two-particle angular correlation function for the SM Background} \label{fig:corr_sm}
\end{subfigure}
\caption{\footnotesize Two-particle angular correlation function, $C^{(2)}(\Delta y, \Delta \phi)$ (in rapidity and azimuthal angle differences) for \subref{fig:corr_hv} pure Hidden Valley signal, and \subref{fig:corr_sm} SM background, in $e^+ e^-$ annihilation generated with PYTHIA8 at centre-of-mass energy $\sqrt{s}=250 \;\mathrm{GeV}$. In the pure Hidden Valley scenario case, we set $m_{q_v}=100 \;\mathrm{GeV}$, $m_{D_v}=125\;\mathrm{GeV}$ and $\alpha_v=0.1$.
For the SM scenario, we consider the production of all quarks up to the bottom $b$.} \label{fig:correlations}
\end{figure} 

We show three-dimensional plots of the two-particle correlation function $C^{(2)}(\Delta y, \Delta \phi)$
separately for HV (Fig.~\ref{fig:corr_hv}) and for SM (Fig.~\ref{fig:corr_sm}) scenarios. The latter compares with the experimentally obtained correlation function displayed by ALEPH from $e^+e^-$ archived data at $\sqrt{s}=206$ GeV~\cite{Badea_2019}. In the HV case, we set $m_{q_v}=100 \;\mathrm{GeV}$, $m_{D_v}=125\;\mathrm{GeV}$ and $\alpha_v=0.1$ and the products of the decay $D_v \to d + q_v$ initiating a partonic (visible and invisible) shower. As expected, a near-side peak shows up at ($\Delta y \simeq 0 $, $\Delta \phi \simeq 0$), receiving contributions mainly from track pairs within the same jet. On the other hand, an away-side correlation ridge, located along $\Delta \phi \simeq \pi$, results from back-to-back momentum balance, in principle unrelated to New Physics (NP).

Notice, however, a clear difference in the structure of the 2-particle correlation function in both 3D plots: a near-side ridge (with two pronounced bumps) shows up for  $1.6 < |\Delta y| < 3$ at $\Delta \phi \simeq 0$ in the pure HV scenario, while it is absent in the SM cascade. We claim that such a different shape of the distribution could be interpreted as a hint of NP, highlighting angular correlations as complementary to more conventional searches beyond the SM. The possibility of
using such a rather diffuse angular signal of NP in multiparticle production at high-energy collisions was put forward in Refs.~\cite{Sanchis-Lozano:2008zjj,Sanchis2009}.

\begin{figure}
\centering
\includegraphics[width=0.55\linewidth]{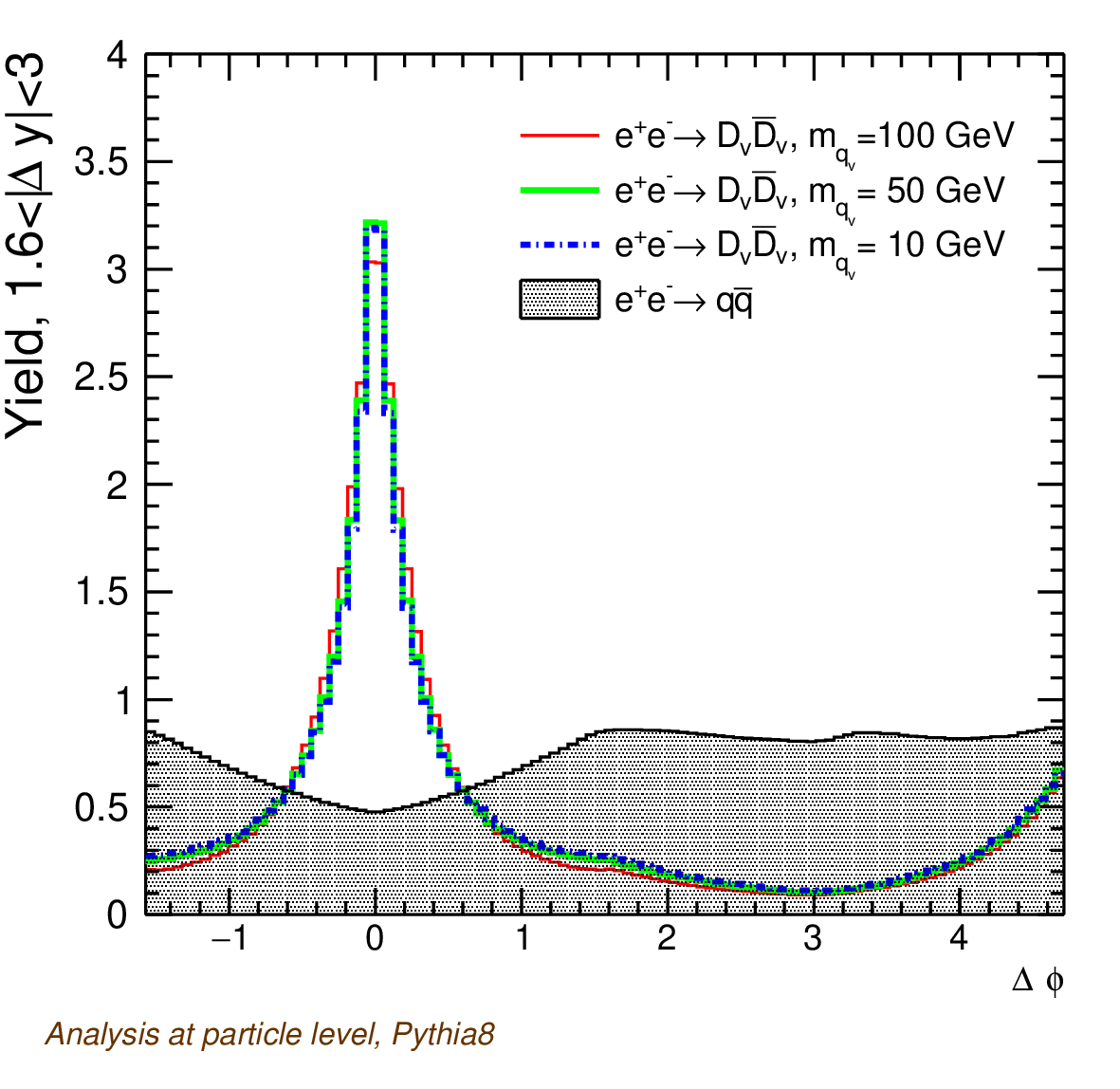}
\caption{\footnotesize \centering Yield Y($\Delta \phi$) for both HV signal (red, green and dashed-blue lines depending on the $q_v$ mass) and SM background (shaded) for the $ 1.6<|\Delta y|<3$ interval. Notice the different shapes of the HV signal and SM background providing a valuable signature of NP using angular correlations.}  \label{fig:yield}

\end{figure}

To examine in more detail the possibility of discriminating the HV signal from the SM background, we depict in 
Figure~\ref{fig:yield} the yield $Y(\Delta \phi)$ (as defined in Eq.~(\ref{eq:yield})) in the $1.6<|\Delta y|<3$ range.  Here the HV signal (for $m_{q_v}$ = 10, 50 and 100 GeV) and the SM background are considered separately, since the sole purpose for the time being is to make a comparison between them at particle level. One can see at once two completely different behaviours
of the yield: a bump at $\Delta \phi \sim 0$ for the HV case while there is a valley in the SM one (as expected). Conversely, 
there is a valley (i.e. no contribution)  at $\Delta \phi \sim \pi$ for the HV case, but not so for the SM one which remains more or less a constant. Such a different behaviour of the HV signal versus the SM background can eventually provide a valuable signature of a hidden sector, which constitutes the main goal of this study.

\FloatBarrier
 
\section{Conclusions}
\label{sec:conclusions}
As is well-known, the analysis of angular particle correlations in high-energy colliders can provide valuable insights into the matter under extreme temperature and density conditions, somehow reproducing the early universe conditions when quarks and gluons had not yet bound to form hadrons.
Furthermore, this approach can also become a complementary tool to uncover the existence of hidden sectors beyond the SM, as stressed in Refs.~\cite{Sanchis-Lozano:2008zjj,Sanchis2009} and postulated in this work.

This study explores the potential for Hidden Valley discovery using two-particle azimuthal correlations at future $e^{+} e^{-}$ colliders. Specifically, we consider a QCD-like scenario in which the SM sector and a hidden one (where new but not too heavy particles, like v-quarks $q_v$ and v-gluons $g_v$, could still remain undiscovered) interact with the SM sector via rather massive (typically $\lesssim 1\;\mathrm{TeV}$ ) communicators. In particular, we have focused on $D_v\bar{D}_v$  pair production, i.e., the lightest communicator being the mirror partner of the SM $d$-quark in this model.

Our preliminary results indicate that, indeed, the analysis of two-particle azimuthal correlations in a $e^+e^-$ factory might become a useful tool to discover New Physics if the centre-of-mass energy is about or larger than twice the communicator mass. Such  searches, based on rather diffuse signals, should be considered complementary to other more conventional searches, thereby increasing the discovery potential of these machines. The extension of the present study to higher energies, e.g. $\sqrt{s}=500\;\mathrm{GeV}-1\;\mathrm{TeV}$, together with the inclusion of detector effects will be addressed in a forthcoming paper.

\section*{Acknowledgements}
AI, EM and VAM acknowledge support by the Generalitat Valenciana via the Excellence Grant 
CIPROM/2021/073 and by the Spanish MICIN/ AEI and the European Union / FEDER via the grant PID2021-122134NB-C21. 
AI also acknowledges support by the Generalitat Valenciana (Spain) under the grant number CIDEGENT/2020/21, the financial support from the MCIN with funding from the European Union NextGenerationEU and Generalitat Valenciana in the call Programa de Planes Complementarios de I+D+i (PRTR 2022) Project \textit{Si4HiggsFactories}, reference ASFAE$/2022/015$.
Imanol Corredoira acknowledges support from Xunta de Galicia (CIGUS Network of Research Centers). M.A.S.L. acknowledges 
support by the Spanish Agencia Estatal de Investigacion under grant PID2020-113334GB-I00 / AEI
/ 10.13039/501100011033, and by Generalitat Valenciana under grant CIPROM/2022/36.


\clearpage
\printbibliography[title=References]

\end{document}